\documentclass{PoS}

\usepackage{amsmath}
\usepackage[caption=false]{subfig}

\title{Lepton angular distributions of Drell-Yan process in pQCD and a geometric approach}

\ShortTitle{Angular distributions of Drell-Yan process}

\author{\speaker{Wen-Chen Chang} \\
  Institute of Physics, Academia Sinica, Taipei 11529, Taiwan}

\author{Randall Evan McClellan \\
Thomas Jefferson National Accelerator Facility, Newport News, VA 23606, USA
}

\author{Jen-Chieh Peng \\
Department of Physics, University of Illinois at Urbana-Champaign, Urbana, Illinois 61801, USA
}

\author{Oleg Teryaev \\
Bogoliubov Laboratory of Theoretical Physics, JINR, 141980 Dubna, Russia
}

\abstract{The lepton angular distributions of the Drell-Yan process in
  the fixed-target experiments are investigated by NLO and NNLO
  perturbative QCD. We present the calculated angular parameters
  $\lambda$, $\mu$, $\nu$ and the degree of violation of the Lam-Tung
  relation, $1-\lambda-2\nu$, for the E615 experiment as well as
  predictions for the COMPASS experiment. Many salient features of
  transverse momentum and rapidity dependence could be qualitatively
  understood by a geometric approach.}

\FullConference{XXVII International Workshop on Deep-Inelastic Scattering and Related Subjects - DIS2019\\
		8-12 April, 2019\\
		Torino, Italy}

\begin{document}

\section{Introduction}

The Drell-Yan (D-Y) process together with deep inelastic scattering
are the main tools for extracting the parton distributions in
hadrons~\cite{peng14}. The polar and azimuthal angular distributions
of leptons produced in unpolarized D-Y process are sensitive to the
underlying reaction mechanisms as well as novel parton
distributions. For example, Boer-Mulders functions~\cite{boer99} have
been suggested to account for a violation of the Lam-Tung (L-T)
relation~\cite{lam80} observed in the fixed-target experiments with
pion beams, e.g. E615~\cite{conway}.

In this proceedings, we compare the data of dilepton angular
parameters $\lambda$, $\mu$, $\nu$ and the L-T violation quantity
$1-\lambda-2\nu$ measured by E615~\cite{conway} with the fixed-order
pQCD calculations~\cite{Lambertsen:2016wgj}. Furthermore we present
the NLO pQCD predictions for the ongoing COMPASS~\cite{COMPASS}
experiments on the dimuon mass $Q$ and Feynman-$x$ ($x_F$) dependence
of these angular parameters. In addition, we interpret some notable
features of pQCD results using the geometric
model~\cite{peng16,chang17,peng18}. More results and greater details
can be found in Ref.~\cite{chang18}.

\section{Calculations of dilepton angular parameters in DYNNLO}

We utilize the DYNNLO (version 1.5) package for the calculations. Via
the LHAPDF6 framework, the parton distribution functions (PDFs) used
for the protons and neutrons are ``CT14nlo'' and ``CT14nnlo'' in the
NLO and NNLO calculations, respectively, and ``GRVPI1'' for the pion
PDFs in both NLO and NNLO calculations.

In the rest frame of the virtual photon in the D-Y process, a commonly
used expression for the lepton angular distributions is given
as
\begin{equation}
\frac{d\sigma}{d\Omega} \propto 1+ \lambda \cos^2\theta
+ \mu \sin 2 \theta\cos\phi
+ \frac{\nu}{2} \sin^2\theta \cos 2 \phi,
\label{eq:eq1}
\end{equation}
where $\theta$ and $\phi$ refer to the polar and azimuthal angles of
$l^-$ ($e^-$ or $\mu^-$). To obtain the $\lambda$, $\mu$, and $\nu$
parameters, we first calculate the $A_i$ parameters in an alternative
expression of the lepton angular distributions of the D-Y process as
follows:
\begin{eqnarray}
\frac{d\sigma}{d\Omega} & \propto & (1+\cos^2\theta)+\frac{A_0}{2}
(1-3\cos^2\theta) +A_1 \sin 2 \theta\cos\phi + \frac{A_2}{2} \sin^2\theta \cos 2 \phi.
\label{eq:eq3}
\end{eqnarray}
The angular coefficients $A_i$ could be evaluated by the moments of
spherical harmonic polynomial expressed as
\begin{eqnarray}
A_0  =  4 - 10 \langle \cos^2\theta \rangle;~~~
A_1  =  5 \langle \sin2\theta \cos \phi \rangle;~~~
A_2  =  10 \langle \sin^2\theta \cos2\phi \rangle,
\label{eq:eq4}
\end{eqnarray}
where $\langle f(\theta, \phi) \rangle$ denotes the moment of
$f(\theta, \phi)$ , i.e. the weighted average of $f(\theta, \phi)$ by
the cross sections in Eq.~(\ref{eq:eq3}). It is straightforward to
show that $\lambda, \mu, \nu$ in Eq.~(\ref{eq:eq1}) are related to
$A_0, A_1, A_2$ via
\begin{eqnarray}
\lambda = \frac{2-3A_0}{2+A_0};~~~ 
\mu  =  \frac{2A_1}{2+A_0};~~~
\nu  =  \frac{2A_2}{2+A_0}.
\label{eq:eq5}
\end{eqnarray}
Equation~(\ref{eq:eq5}) shows that the L-T relation, $1-\lambda - 2
\nu=0$, is equivalent to $A_0 = A_2$.

\begin{figure}[htbp]
\centering
\includegraphics[width=0.8\columnwidth]{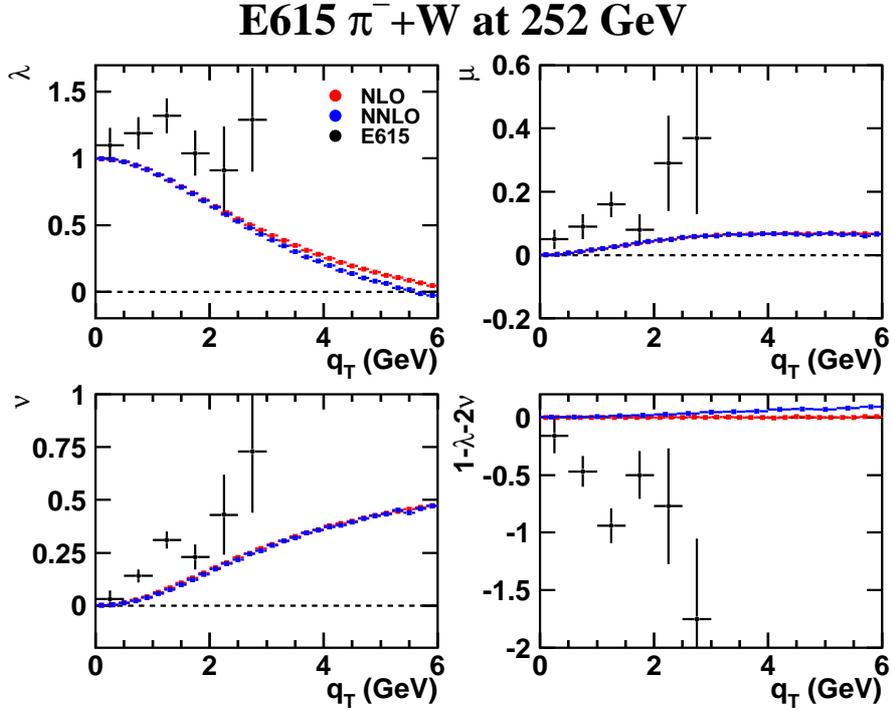}
\caption{Comparison of NLO (red points) and NNLO (blue points)
  fixed-order pQCD calculations with the E615 $\pi^-+W$ D-Y data at 252
  GeV~\cite{conway} (black points) for $\lambda$, $\mu$, $\nu$ and
  $1-\lambda-2\nu$.}
\label{fig1_e615}
\end{figure}

In Fig.~\ref{fig1_e615}, we compare the results of $\lambda$, $\mu$,
$\nu$, and the L-T violation, $1-\lambda-2\nu$, from the fixed-order
pQCD calculations with 252-GeV $\pi^- + W$ data from E615
experiment~\cite{conway}. The angular parameters are evaluated as a
function of the dimuon's transverse momentum ($q_T$) in the
Collins-Soper frame. Overall, the calculated $\lambda$, $\mu$ and
$\nu$ exhibit distinct $q_T$ dependencies. At $q_T \rightarrow 0$,
$\lambda$, $\mu$ and $\nu$ approach the values predicted by the
collinear parton model: $\lambda = 1$ and $\mu = \nu =0$. As $q_T$
increases, Fig.~\ref{fig1_e615} shows that $\lambda$ decreases toward
its large-$q_T$ limit of $-1/3$ while $\nu$ increases toward $2/3$,
for both $q\bar{q}$ and $qG$ processes~\cite{thews,lindfors}. The
$q_T$ dependence of $\mu$ is relatively mild compared to $\lambda$ and
$\nu$. This is understood as a result of some cancellation effect, to
be discussed in Sec.~\ref{sec:discussion}.

Comparing the results of the NLO with the NNLO calculations, $\lambda
{\rm (NNLO)}$ is smaller than $\lambda \rm{(NLO)}$ while $\mu$ and
$\nu$ are very similar for NLO and NNLO. The amount of L-T violation,
$1-\lambda-2\nu$, is zero in the NLO calculation, and nonzero and
positive in the NNLO calculation. As seen in Fig.~\ref{fig1_e615}, the
pQCD predicts a sizable magnitude for $\nu$, comparable to the
data. Therefore, the pQCD effect should be included in the extraction
of nonperturbative Boer-Mulders effect from the data of $\nu$.

Next we present the results of the angular coefficients $\lambda$,
$\mu$ and $\nu$ as a function of $q_T$ in various bins of $Q$ and
$x_F$ for the ongoing fixed-target COMPASS experiment at
CERN~\cite{COMPASS}. This experiment runs with 190-GeV $\pi^-$ beam
and transversely-polarized $\rm{NH}_3$ target and unpolarized aluminum
($Al$) and tungsten ($W$) nuclear targets. There are three bins for
$Q$ in the range of 4.0--7.0 GeV, as well as three bins for $x_F$ in
the range of 0--0.6. Our results could be convoluted by the COMPASS
spectrometer acceptances later for a direct comparison with
experimental data in the near future. Since there are no significant
difference between the NLO and NNLO results, we present only the
results from the NLO calculation to illustrate the major features.

\begin{figure}[htbp]
\centering
\subfloat[]
{\includegraphics[width=0.8\columnwidth]{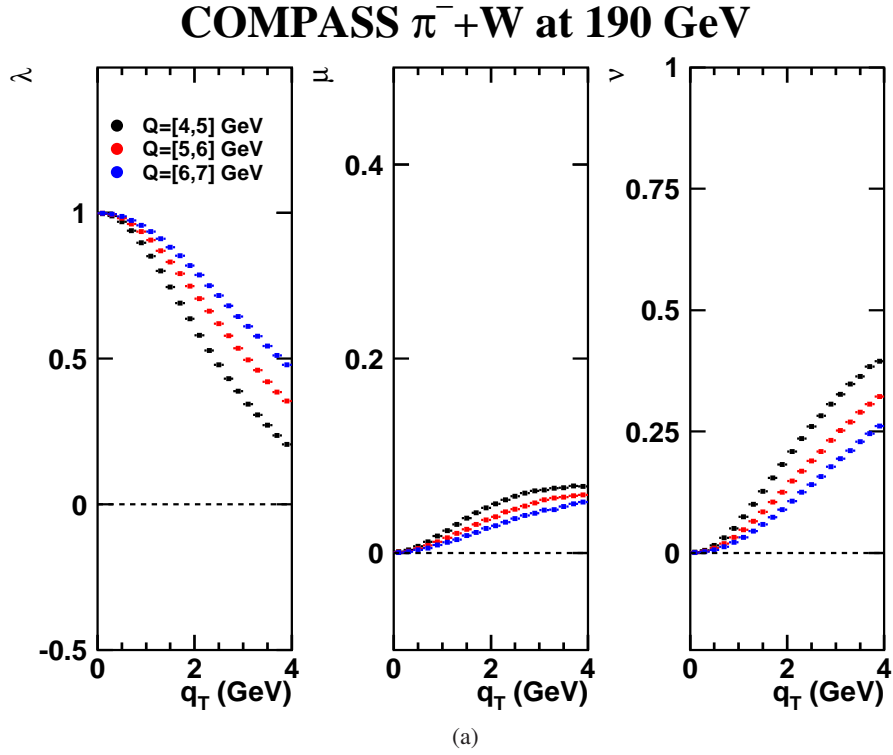}}\\
\subfloat[]
{\includegraphics[width=0.8\columnwidth]{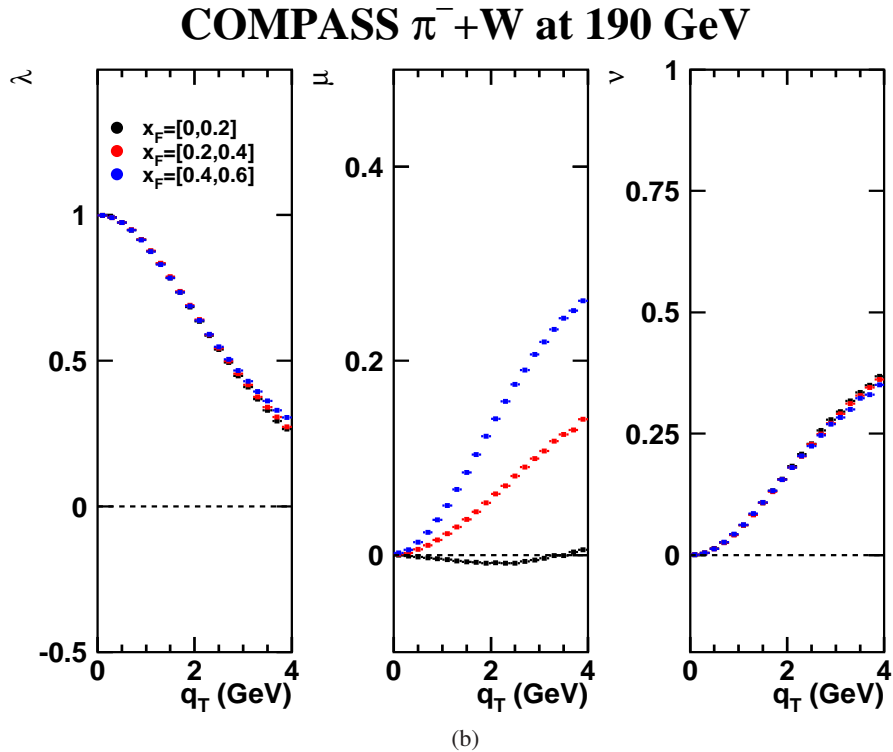}}
\caption
[\protect{}] {(a) NLO pQCD results of $\lambda$, $\mu$, and $\nu$ as a
  function of $q_T$ at several dimuon mass $Q$ bins and $x_F>0$ for
  D-Y production off the tungsten target with 190-GeV $\pi^-$ beam in
  the COMPASS experiment. (b) Same as (a) but at several Feynman-$x$
  $x_F$ bins and $4<Q<9$ GeV.}
\label{fig2_compass}
\end{figure}

Figures~\ref{fig2_compass}(a) and (b) show $\lambda$, $\mu$ and $\nu$
as a function of $q_T$ for various bins of the dimuon mass, $Q$, and
Feynman-$x$, $x_F$, respectively. The $q_T$ distributions of $\lambda$
and $\nu$ parameters depend sensitively on $Q$, but only weakly on
$x_F$. As for $\mu$, its $q_T$ distribution has strong dependencies on
$x_F$ and on $Q$. In particular, the magnitude of $\mu$ is small when
$x_F$ is close to 0 and its sign could even turn negative at some
$q_T$ region. As $x_F$ increases, the magnitude of $\mu$ increases
pronouncedly.

\section{Geometric model}
\label{sec:discussion}

The E615 D-Y data of lepton angular distributions can be reasonably
well described by the NLO and NNLO pQCD calculations. Various salient
features of $Q$ and $x_F$ dependencies are observed in the predicted
results of $\lambda$, $\mu$ and $\nu$ parameters for COMPASS
experiment based on NLO pQCD. It is of interest to check if these
features of pQCD calculations could be understood using the geometric
approach developed in Refs.~\cite{peng16,chang17}.

In Refs.~\cite{peng16,chang17}, the hadron plane, the quark plane, and
the lepton plane of collision geometry are defined in the
Collins-Soper $\gamma^*$ rest frame. A pair of collinear $q$ and $\bar
q$ with equal momenta annihilate into a $\gamma^*$. The momentum unit
vector of $q$ is defined as $\hat z^\prime$, and the quark plane is
formed by the $\hat z^\prime$ and the $\hat z$ axes of Collins-Soper
frame. The angular coefficients $A_i$ in Eq.~(\ref{eq:eq3}) can be
expressed in term of $\theta_1$ and $\phi_1$ as follows:
\begin{eqnarray}
A_0 =  \langle\sin^2\theta_1\rangle, ~~~
A_1 = \frac{1}{2} \langle\sin 2\theta_1\cos \phi_1\rangle, ~~~
A_2 =  \langle\sin^2\theta_1 \cos 2\phi_1\rangle,
\label{eq:eq8}
\end{eqnarray}
where the $\theta_1$ and $\phi_1$ are the polar and azimuthal angles
of the natural quark axis $\hat z^\prime$ of the quark plane in the
Collins-Soper frame. The $\langle \cdot \cdot \cdot \rangle$ in
Eq.~(\ref{eq:eq8}) denotes that the measured values of $A_i$ at a
given kinematic bin are averaged over events having particular values
of $\theta_1$ and $\phi_1$.

Up to NLO ($\mathcal{O}(\alpha_S)$) in pQCD, the quark plane coincides
with the hadron plane and $\phi_1=0$. Therefore $A_0=A_2$ or
$1-\lambda-2\nu=0$, i.e., the L-T relation is satisfied. Higher order
pQCD processes allow the quark plane to deviate from the hadron plane,
i.e., $\phi_1 \neq 0$. This acoplanarity effect leads to the violation
of the L-T relation. For a nonzero $\phi_1$, Eq.~(\ref{eq:eq8}) shows
that $A_2 < A_0$. Therefore, when the L-T relation is violated, $A_0$
must be greater than $A_2$ or, equivalently, $1 - \lambda - 2\nu
>0$. This expectation of $1 - \lambda - 2\nu >0$ in the geometric
approach agrees with the results of NNLO pQCD calculations shown in
Fig.~\ref{fig1_e615}. The geometric approach offers a simple
interpretation for this result.

Furthermore the sign of $\mu$ could be either positive or negative,
depending on which parton and hadron the gluon is emitted
from~\cite{chang17,chang18}. Hence, one expects some cancellation
effects for $\mu$ among contributions from various processes. Each
process is weighted by the corresponding density distributions for the
interacting partons. At $x_F \sim 0$, the momentum fraction carried by
the beam parton ($x_B$) is comparable to that of the target parton
($x_T$). Therefore, the weighting factors for various processes are of
similar magnitude and the cancellation effect could be very
significant, resulting in a small value of $\mu$. As $x_F$ increases
toward 1, $x_B$ becomes much larger than $x_T$. In this case the
weighting factors are now dominated by fewer processes, resulting in
less cancellation and a larger value of $\mu$. This explains why the
$\mu$ parameter exhibits a strong $x_F$ dependence in
Figs.~\ref{fig2_compass}(b).

\section {Summary}

We have presented a comparison of the measurements of the angular
parameters $\lambda$, $\mu$, $\nu$ and $1-\lambda-2\nu$ of the D-Y
process from the fixed-target E615 experiment with the corresponding
results from the NLO and NNLO pQCD calculations. Qualitatively the
transverse momentum ($q_T$) dependence of $\lambda$, $\mu$ and $\nu$
in the data could be described by pQCD. The difference between NLO and
NNLO results becomes visible at large $q_T$. The L-T violation part
$1-\lambda-2\nu$ remains zero in the NLO pQCD calculation and turns
positive in NNLO pQCD.

The $x_F$ dependence of the angular parameters is well described by
the geometric picture. In particular, the weak rapidity dependencies
of the $\lambda$ and $\nu$, and the pronounced rapidity dependency for
$\mu$ can be explained by the absence or presence of
rapidity-dependent cancellation effects. The occurrence of
acoplanarity between the quark plane and the hadron plane ($\phi_1
\neq 0$), for the pQCD processes beyond NLO leads to a violation of
the L-T relation. The predicted positive value of $1-\lambda-2\nu$, or
$A_0>A_2$ when $\phi_1$ is nonzero, is consistent with the NNLO pQCD
results.

\end{document}